\documentclass{aa}
\input{epsf.sty}
\usepackage{graphicx}
\def\beq{\begin{equation}}
\def\enq{\end{equation}}

\begin{document}
%\thesaurus{09.03.2, 09.09.1, 13.07.2}

\title{SPI observations of the diffuse $^{60}$Fe emission in the Galaxy}

\institute{Max-Planck-Institut f\"{u}r extraterrestrische Physik,
Postfach 1603, 85740 Garching, Germany \and AstroParticule et
Cosmologie (APC), 11 Place Marcelin Berthelot, 75231 Paris, France
\and Centre d'$\acute{\rm E}$tude Spatiale des Rayonnements, B.P.
4346, 31028 Toulouse Cedex 4, France \and DSM/DAPNIA/SAp, CEA
Saclay, 91191 Gif-sur-Yvette, France \and Space Sciences
Laboratory, UC Berkeley, CA 94720, USA }

\author{W.~Wang\inst{1}, M.J.~Harris\inst{1}, R.~Diehl\inst{1},
H.~Halloin\inst{2}, B. Cordier\inst{4}, A.W. Strong\inst{1}, K.
Kretschmer\inst{1}, J. Kn\"odlseder\inst{3}, P. Jean\inst{3}, G.G.
Lichti\inst{1}, J.P. Roques\inst{3}, S. Schanne\inst{4}, A. von
Kienlin\inst{1}, G. Weidenspointner\inst{3}, C. Wunderer\inst{5}}

\offprints{W. Wang, wwang@mpe.mpg.de}
\date{Received }

\authorrunning{W. Wang et al.}
\titlerunning{SPI observations of $^{60}$Fe}

\abstract
{{\it Aims.} Gamma-ray line emission from radioactive decay of
$^{60}$Fe provides constraints on nucleosynthesis in massive stars
and supernovae.

{\it Methods.} The spectrometer SPI on board {\em INTEGRAL\/} has
accumulated nearly three years of data on gamma-ray emission from
the Galactic plane. We have analyzed these data with suitable
instrumental-background models and sky distributions to produce
high-resolution spectra of Galactic emission.

{\it Results.} We detect the $\gamma$-ray lines from $^{60}$Fe
decay at 1173 and 1333~keV, obtaining an improvement over our
earlier measurement of both lines with now 4.9~$\sigma$
significance for the combination of the two lines. The average
flux per line is $ (4.4 \pm 0.9) \times 10^{-5} ~{\rm ph}
~\hbox{cm}^{-2} \hbox{s}^{-1}~\hbox{rad}^{-1}$ for the inner
Galaxy region. Deriving the Galactic $^{26}$Al gamma-ray line flux
with using the same set of observations and analysis method, we
determine the flux ratio of $^{60}{\rm Fe}/^{26}{\rm Al}$
gamma-rays as $0.148 \pm 0.06$.

{\it Conclusions} The current theoretical predictions are
consistent with our result. We discuss the implications of these
results for the widely-held hypothesis that $^{60}$Fe is
synthesized in core-collapse supernovae, and also for the
closely-related question of the precise origin of $^{26}$Al in
massive stars.

 }

\maketitle

\section{Introduction}

The radioactive isotope $^{60}$Fe is believed to be synthesized
through successive neutron captures on Fe isotopes (e.g.
$^{56}$Fe) in a neutron-rich environment inside massive stars,
before or during their final evolution to core collapse supernovae
(CCSN). Due to its long decay time ($\tau \simeq$2.2~My),
$^{60}$Fe survives to be detected after the supernova ejected it
into the interstellar medium, by $\beta$-decay via $^{60}$Co and
$\gamma$ emission at $1173$ keV and 1333 keV -- like other
radioactive isotopes: $^{44}$Ti, $^{56,57}$Co, and $^{26}$Al.
These isotopes provide evidence that nucleosynthesis is ongoing in
the Galaxy (with CCSN occurring at intervals $\sim 30 -100$ yr,
The et al. 2006; Diehl et al. 2006). The discoveries of the above
isotopes have fulfilled predictions of nucleosynthesis theory, at
least in its general outline.  The $^{44}$Ti and $^{56,57}$Co
isotopes have relatively short half-lives ($\leq$ 100~years) and
are thus detected as point-like sources, e.g. in young Type II
supernova remnants (SNRs) Cas A and 1987A, respectively (Iyudin et
al. 1994, 1995; Matz et al. 1988; Kurfess et al. 1992).  With its
much longer half-life ($\sim 10^{6}$ years), $^{26}$Al may
propagate over significant distances of $\simeq$~few hundred pc,
and accumulates in the interstellar medium (ISM) from many
supernovae, until injection and $\beta$-decay are in balance in
the ISM, giving rise to a diffuse and even Galaxy-wide glow
(Mahoney et al. 1982; Prantzos and Diehl 1996). The behavior of
$^{60}$Fe from the same massive-star sources should follow that of
$^{26}$Al, since its half-life is similar, $\sim 2.2 \times
10^{6}$ years. In addition, $^{60}$Fe could also be produced in
substantial amounts by rare subtypes of SN Ia (Woosley 1997),
which would then be point sources of $^{60}$Fe gamma-rays.

Although the detections of these isotopes are in agreement with
the broad outlines of CCSN nucleosynthesis theory, there are
discrepancies in the details. For example, $^{44}$Ti lives long
enough so that it should have been detected from several recent
Galactic supernovae if these occur at a rate of $\sim 2$ per
century; but with the exception of Cas A (Iyudin et al. 1994,
Renaud et al. 2004, Renaud et al. 2006), no other SNRs have been
detected yet. Mapping of the diffuse $^{26}$Al emission
(Kn\"{o}dlseder et al. 1999, Pl\"uschke et al. 2001) confirms that
it follows the overall Galactic massive star population, but
concentrations of $^{26}$Al are seen in very young OB associations
where even the most massive stars are not expected to have
exploded (e.g. Cygnus OB associations). This has led to the
suggestion that most $^{26}$Al production is associated with
ejections in an earlier, pre-explosion phase of stellar evolution
(Kn\"odlseder et al. 1999).

Measurements of $^{60}$Fe promise to provide new information about
these issues, specifically the massive star nucleosynthesis in the
late pre-supernova stages. RHESSI reported observations of the
gamma-ray lines from $^{60}$Fe with 2.6 $\sigma$ significance, and
an average flux of $ (3.6\pm 1.4) \times 10^{-5} {\rm ph\ cm^{-2}\
s^{-1}}$ (Smith 2004). The analysis of the first year of data from
the SPI spectrometer on ESA's {\em INTEGRAL} spacecraft resulted
in a similarly marginally-significant detection of these
$\gamma$-ray lines from $^{60}$Fe ($\sim 3 \sigma$, Harris et al.
2005), with an average line flux of $(3.7 \pm\ 1.1) \times 10^{-5}
{\rm ph\ cm^{-2}\ s^{-1}}$. SPI on {\em INTEGRAL\/} has
accumulated more data since then; here we analyzed 2.5 years of
data, aiming at a consolidation of the {\em INTEGRAL}/SPI
measurement of $^{60}$Fe gamma-rays.

In the following section, we describe our observations with SPI.
In \S 3, we describe the steps and methods of data analysis.
The scientific results, including the spectra and intensities of
both 1173 keV and 1333 keV lines, and the flux ratio of
$^{60}$Fe/$^{26}$Al are presented in \S 4. Discussions and a
summary follow in \S 5.

\section{Observations}

The {\em INTEGRAL\/} spacecraft was launched on October 17, 2002,
into a high-inclination, high-eccentricity orbit intended to avoid
the increased background from the Earth's trapped radiation belts.
{\em INTEGRAL}'s orbital period is 3 days. The spectrometer SPI
consists of 19 Ge detectors actively shielded by a BGO
anti-coincidence shield. It has a tungsten coded mask in its
aperture which allows imaging at $\sim 3^{\circ}$ resolution
within a $16^{\circ} \times 16^{\circ}$ full coded field of view
(imaging on {\em INTEGRAL\/} is mainly performed at lower energies
by the IBIS telescope, with which SPI is co-aligned). The Ge
detectors are sensitive to gamma-rays between 15 keV and 8 MeV,
with a total effective area $\sim 70$ cm$^{2}$ at 1 MeV, and
achieve an energy resolution of $\sim 2.5$ keV at 1 MeV (Roques et
al. 2003, Atti\'{e} et al. 2003). However, cosmic-ray (CR) impacts
degrade this resolution over time, and the instrument is
periodically shut off for a few days while annealing (by heating
from cryogenic temperatures to 100$^{\circ}$C) is applied to the
detectors to restore the energy resolution by thermal curing
heating of the CR-induced defects (Roques et al. 2003, Leleux et
al. 2003).

In space operations, {\em INTEGRAL} with its IBIS and SPI
telescopes is pointed at predesignated targets, with a fixed
orientation for intervals of typically $\sim 2000$ s (referred to
as {\em pointings\/}), which are successively arranged as a
standard pattern of neighbouring pointings $\simeq 2^\circ$ apart
({\em dithering\/}), and covering target region of interest for
improved imaging (Courvoisier et al. 2003).

The interval covered by the observations analyzed here is December
3, 2002 --- August 22, 2005 (orbits 17--359).  The instrument was
in operation for most of this time, but important gaps were caused
by five of the annealing episodes described above, and by the
regular perigee passages with gaps due to the radiation belts.
The live time was further reduced by the failure of two of the 19
detectors during this interval (December 2003 and July 2004) and
by the occurrences of solar flares.

\section{Data analysis}

The basic measurement of SPI consists of event messages per photon
triggering the Ge detector camera. We distinguish events which
trigger a single Ge detector element only (hereafter {\it single
event}, SE), and events which trigger two Ge detector elements
nearly simultaneously (hereafter {\it multiple event}, ME), which
is the case for $\simeq 30\%$ of $^{60}$Fe gamma-ray photons.

The analysis steps are: (1) Assembly of data selected to be free
of contamination by, e.g., solar-flare events. (2) Modelling the
instrumental background. (3) Fitting the measured data in narrow
energy bins with the background model and a model of celestial
gamma-ray emission, folded through the instrumental response into
the data space of the measurement; the fitted amplitude of the
celestial model per energy bin then comprises the spectrum of
observed sky emission. (4) Deriving line parameters of the
celestial signal. From the two event types, and from the two
gamma-ray lines of $^{60}$Fe, we obtain four independent
measurements of $^{60}$Fe emission.

In the following, we discuss the analysis steps and their
implementation in the corresponding software utilities within
MPE's {\em INTEGRAL} data analysis system.

\subsection{Data selection and assembly}
We test data quality per each `science window' (typically a time
interval of $\sim 30$ minutes corresponding to one pointing),
applying selection limits to `science housekeeping' parameters
such as the count rates in several onboard radiation detectors,
instrument status codes, data ownership, and orbit phase (utility
$spiselectscw$). We employ the {\em INTEGRAL} Radiation
Environment Monitor (IREM; Hajdas et al. 2003), the SPI plastic
scintillator anticoincidence counter (PSAC), and the rate of
saturating events in SPI's Ge detectors (from events depositing
$>$ 8 MeV in a detector; hereafter referred to as GEDSAT rates) to
exclude solar-flare events and other erratic background increases.
Regular perigee background increases are additionally eliminated
through a 0.05--0.95 window on orbital phase. From the selected
events, spectra are accumulated per each detector and pointing,
and together with dead time and pointing information assembled
into the analysis database.

We establish databases of spectra per pointing per detector for
single and multiple events (SE and ME), in spectral ranges with
1~keV bin size in $\sim 20$~keV bands around each of the $^{60}$Fe
lines, and in adjacent energy bands for determination of
instrumental background (see below). Typical such raw data spectra
around two $\gamma$-ray lines are presented in Figure 1, and show
that instrumental background including strong instrumental lines
dominates the raw spectra. The energy ranges around the two lines
extend from 1153 - 1193 keV and 1313 - 1359 keV. For the
background modelling from the adjacent continuum, we use 1163 -
1169 keV plus 1177 - 1184 keV, and 1318 - 1328 keV plus 1336 -
1349 keV energy bands, respectively. From orbits 17 to 359, a
total observation time of $\sim 24$ Ms is thus obtained, with a
data set consisting of 14623 pointings with one spectrum per each
detector and event type.

\begin{figure*}
\centering
\includegraphics[width=7cm]{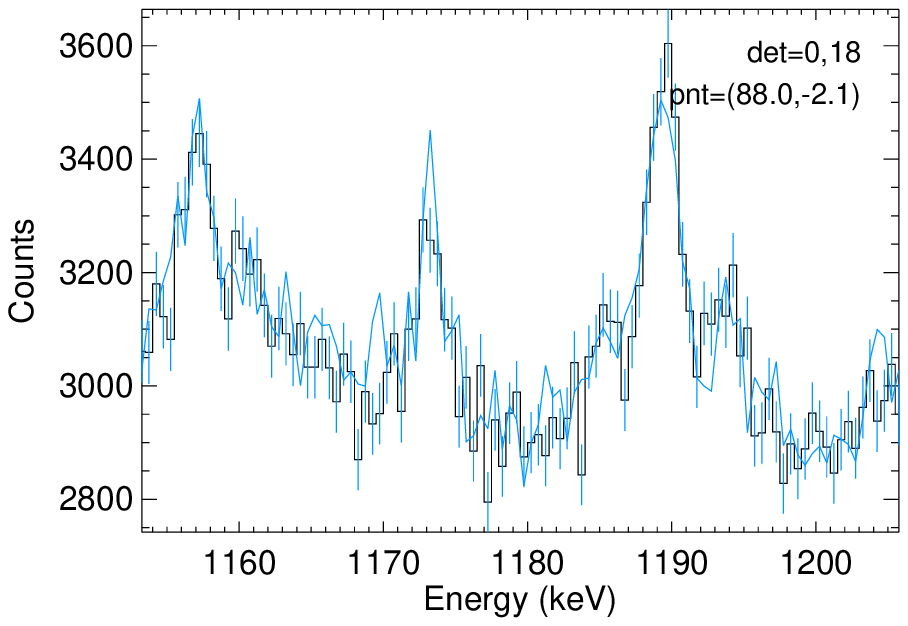}
\includegraphics[width=7cm]{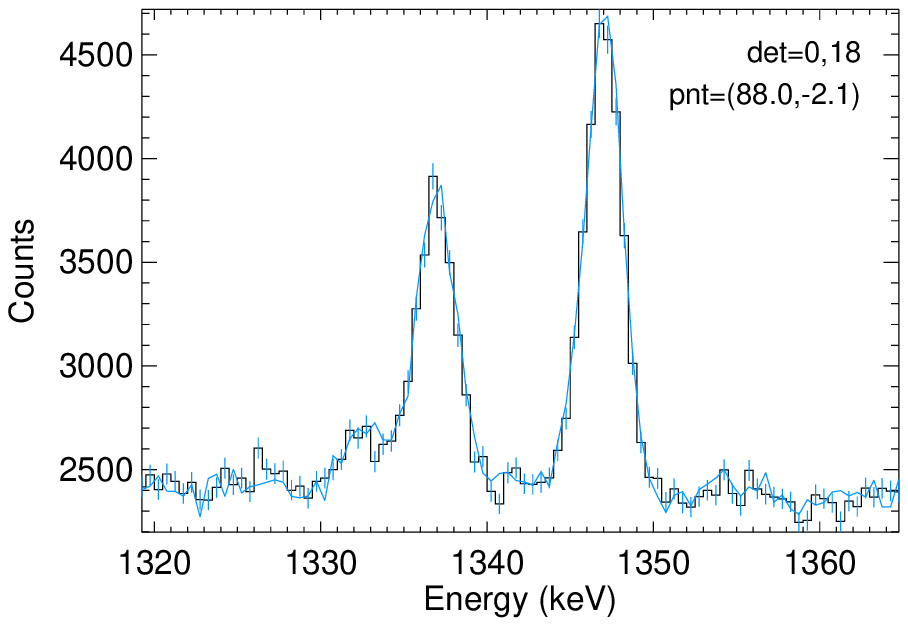}
\caption{Raw data spectra around the energies of two $^{60}$Fe
lines in one-INTEGRAL-orbit observations (3 days), representing
the instrumental lines and continuum background. For the 1173 keV
case (left), three strong instrumental lines are obvious:
$^{44}$Sc (1157 keV), $^{60}$Co (1172.9 keV), $^{182}$Ta (189.4
keV), and the $^{60}$Co line blends with the $^{60}$Fe line
(Weidenspointner et al. 2003). For the 1333 keV case (right), the
$^{60}$Co background line (1332.5 keV) and the other strong
instrumental line of $^{69}$Ge (1336.8 keV) blend with the
$^{60}$Fe line, an instrumental line at 1347 keV also comes from
$^{69}$Ge electron captures. }

\end{figure*}

\begin{figure*}
\centering
\includegraphics[width=6.5cm]{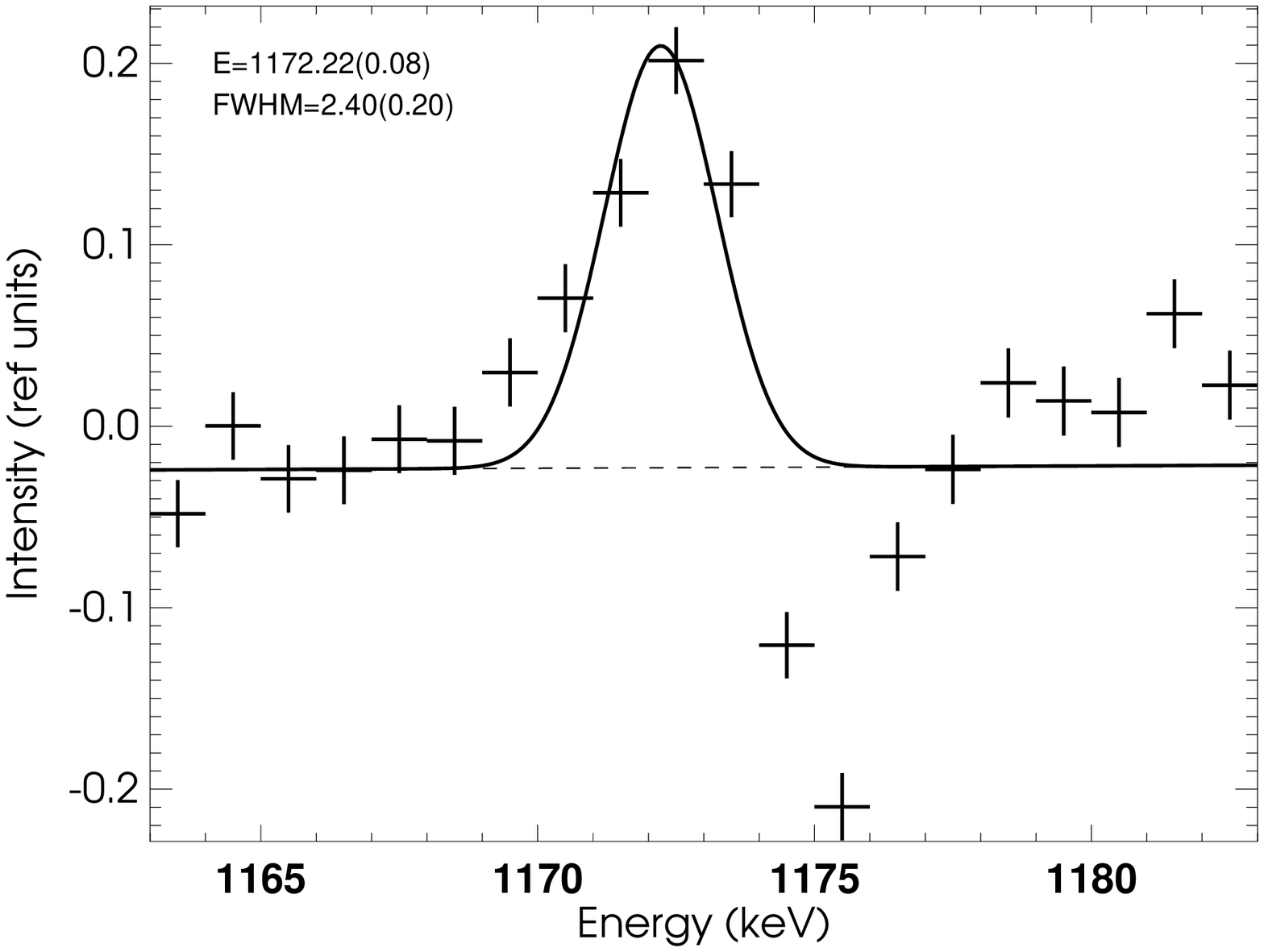}
\includegraphics[width=6.5cm]{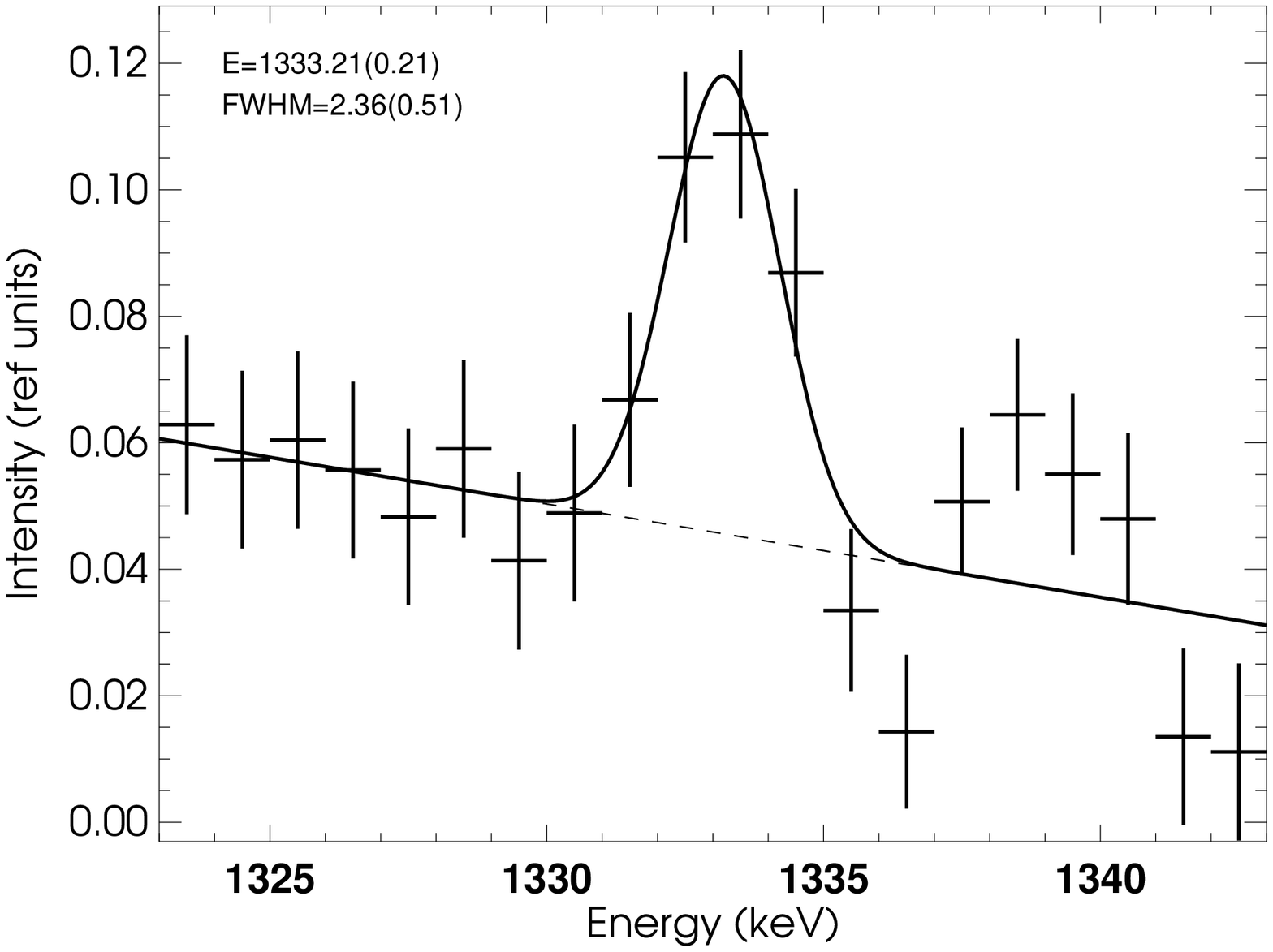}
\caption{Background coefficients for the $^{60}$Co radioactivity
build-up model around the 1173 keV (left) and 1333 keV (right)
lines, respectively.  }
\end{figure*}

\begin{figure*}
\centering
\includegraphics[width=8.5cm]{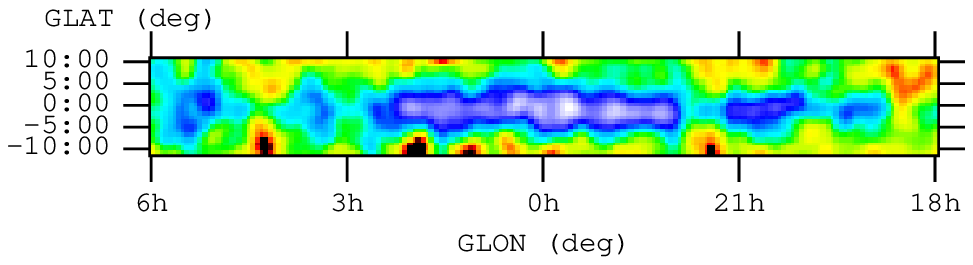}
\includegraphics[width=8.5cm]{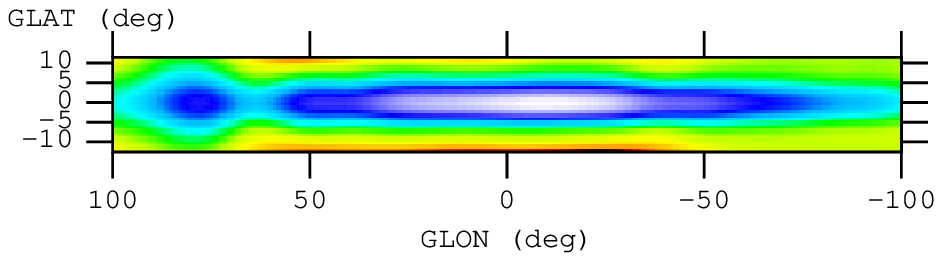}
\caption{Two plausible sky distribution models adopted for diffuse
$^{60}$Fe emission from the Galaxy. Both have been derived from
COMPTEL $^{26}$Al measurements (Pl\"uschke et al., 2001;
Kn\"odlseder et al., 1999). The image obtained with the
Maximum-entropy method (left) shows more structure, while the
image derived with the Multi-resolution Expectation Maximization
Method (MREM) appears smooth from its noise-filtering, accepting
only image structure which is enforced by the measurements. }
\end{figure*}

%%%%%%%%%%%%%%%%%%%%%%%%%%%%%%%%%%%%%%%%%%%%%%%%%%%%%%%%%%%%%%%%%%%%%%%%%%%%%%%
\begin{figure*}
\centering
\includegraphics[width=10.5cm]{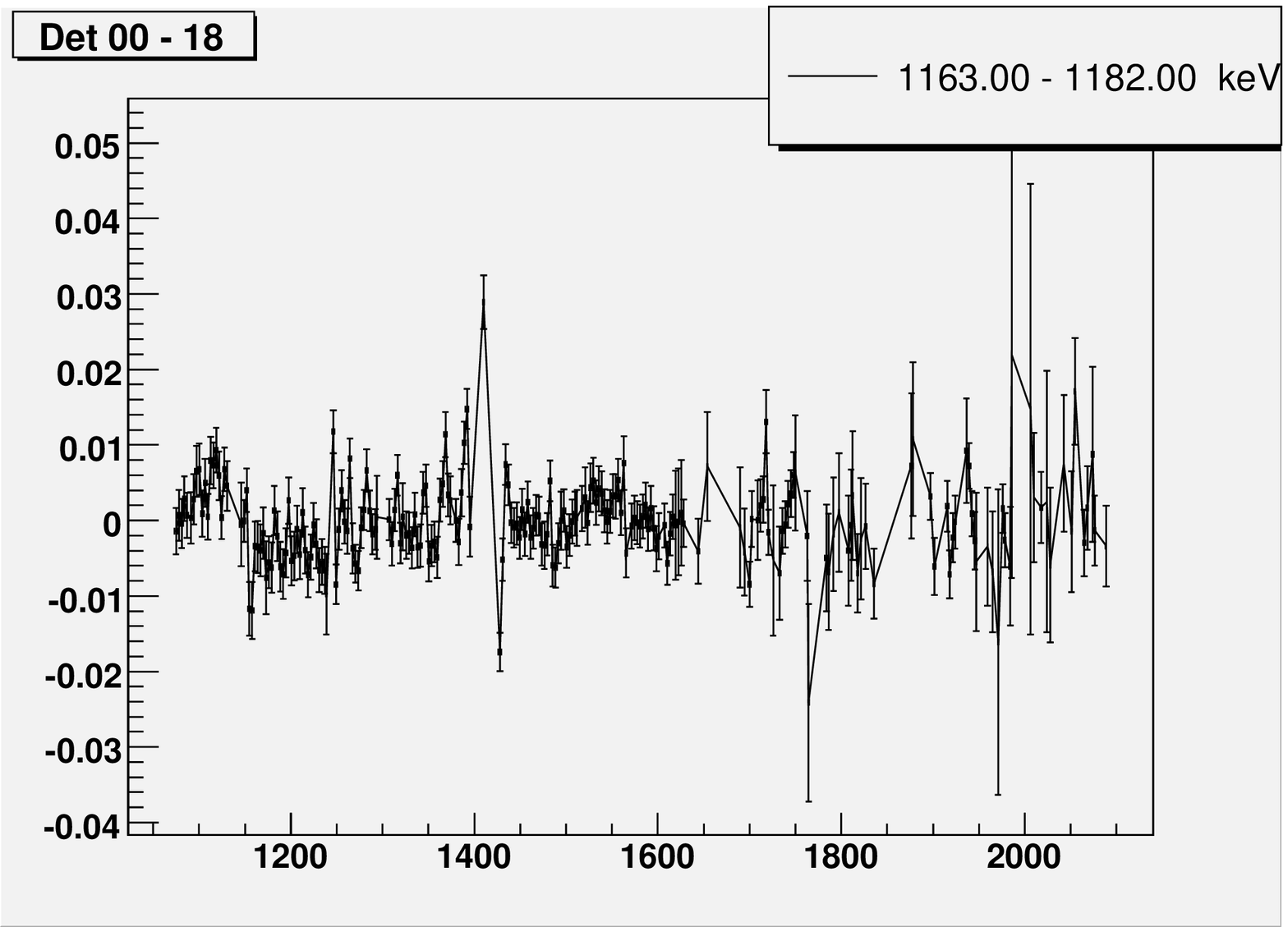}
\includegraphics[width=8.0cm]{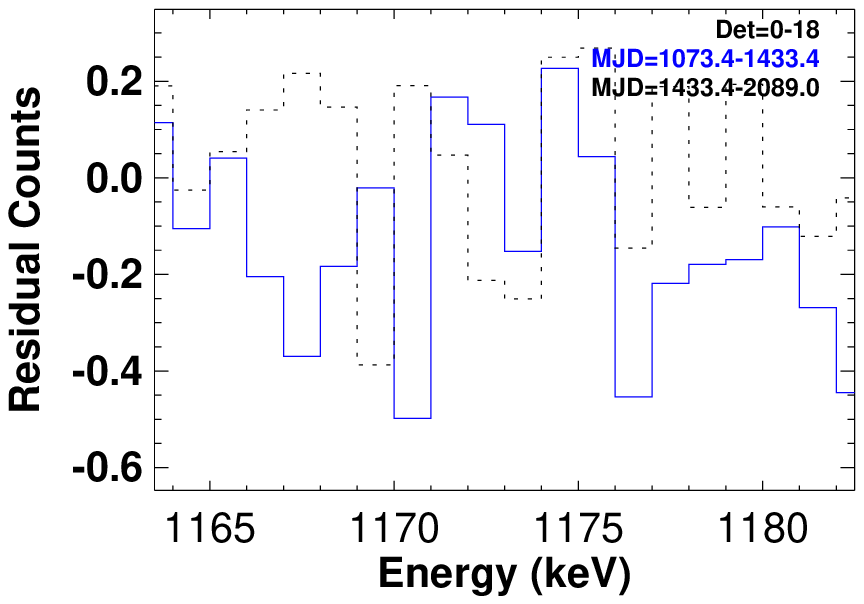}
\includegraphics[width=8.0cm]{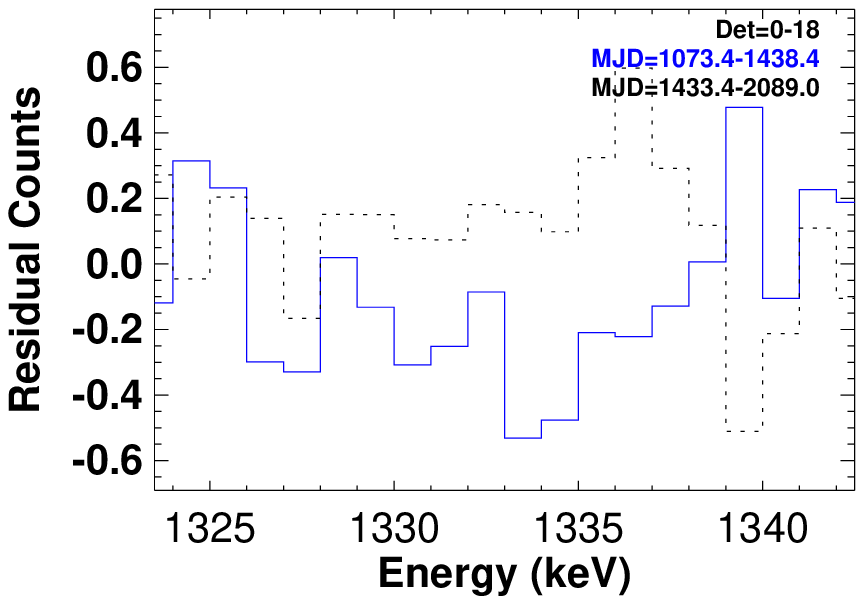}
\caption{Residuals of counts after fitting, in different
projections. The upper figure shows the residuals versus time
(days of the Julian Date, starts at 1 Jan 2000), where counts per
pointing have been re-binned into 3-day intervals for clarity
(here for single events SE in the 1163
--- 1182 keV energy band). The lower two figures present the
residuals projected with the energy (SE cases for the bands around
the 1173 and 1333 keV lines separately). Two different time
periods are shown (dashed: first half of database; solid: second
half of database).  For the 1333 keV line case, because of the
strong instrumental $^{69}$Ge line at 1337 keV, there still exist
the large residuals around 1337 keV. Residuals around zero
generally confirm that our background models are adequate except
for the instrumental line around 1337 keV, with $\chi^2/d.o.f.$ in
model fittings (\S 3.3): 1.182 (SE) and 0.670 (ME) for the 1173
keV line; 1.191 (SE) and 0.667 (ME) for the 1333 keV line. }
\end{figure*}
%%%%%%%%%%%%%%%%%%%%%%%%%%%%%%%%%%%%%%%%%%%%%%%%%%%%%%%%%%%%%%%%%%%%%%%%%%%%%%%

\subsection{Background modelling}

SPI spectra are dominated by the intense background radiation
characteristic of space platforms undergoing CR bombardment.  Much
of this radiation is prompt, resulting directly from CR impacts,
whose variation with time ought to follow that of the incident CR
flux.  Other components arise from radioactive isotopes produced
by the CR impacts, whose decay lifetimes are long compared to the
coincidence window of the detector-anticoincidence electronics.
Radioactive isotopes will increase in abundance until CR
production balances $\beta$-decay.  A time series of their
$\gamma$-ray emission will be the convolution of the time behavior
of the prompt CR source with an exponential decrease from decay.
Local radioactivity in the spacecraft and instruments themselves
thus will generate both broad continuum background emission and
narrow gamma-ray lines from long-lived radio-isotopes. Varying
with energy, background components may exhibit complex time
variability due to their origins from more than one physical
source.

We derive models for the background contribution per energy bin
and pointing in each detector from comparisons to presumed
`tracers' of background; for these, we use independent
contemporaneous measurements aboard {\em INTEGRAL} (see Halloin et
al. 2007, Diehl et al. 2006, also background model studies by
Sizun et al. 2007). Comparing different candidate tracers to
observed background variations, we find the best tracer(s) of
background for each energy band. Candidate tracers with high
statistical precision are the rates measured in SPI's plastic
scintillator anticoincidence detector, the rate of saturated
events in the BGO shield, the detector-by-detector GEDSAT rates
(from events depositing $>$8 MeV in a Ge detector), and
Ge-detector rates integrated over rather wide energy bands (for
better statistical precision). For our analysis, we make use of
the GEDSAT rates tracing the prompt CR activation to the $^{60}$Fe
lines (see above).

Each such tracer is regarded as a `template' for time variability
of background. By normalizing such a template to the set of counts
per detector spectrum (i.e. per pointing), a background model can be
constructed, which is applicable to the analysis data set of spectra.
We construct our background model from two components, one derived
from adjacent energy bands, and an additional component modelling
instrumental line contributions.

The most important radioactive background line component for our
analysis is the emission from $^{60}$Co decay inside the
instrument, whose two lines at 1173 and 1333 keV are actually part
of the celestial $^{60}$Fe decay chain we want to measure. The
convolution of the GEDSAT CR source tracer with an exponential
function $e^{-t/\tau}$, in which the decay time $\tau$ is that of
$^{60}$Co (7.6 yr), is expected to give a good background model
for energy bins containing contributions from these lines, i.e.
around 1173 and 1333 keV (see Figure 2).  Another possible
radioactive contaminant in our spectral region of interest may be
the strong $^{69}$Ge K-shell electron capture line (1337 keV,
Weidenspointner et al. 2003) which blends into the 1333 keV
$^{60}$Fe line; its lifetime is 2.35 days. Because of the short
decay time is short, we just take the GEDSAT as a background
component tracer for this strong instrumental line feature.

Thus, in a first step, the detector-by-detector count rates in the
four continuum bands are fitted by the GEDSAT time series to
construct an `adjacent-energies' background template. Then, the
set of detector-by-detector spectra per pointing in 1 keV bins
covering the $\sim 20$ keV intervals around and including the
$^{60}$Fe lines are fitted to the sum of adjacent continuum
template, plus the radioactivity template for $^{60}$Co, plus an
additional GEDSAT template to capture any additional prompt
background. Thus we generate a background model for our actual set
of spectra, which is constrained in its variability by the various
background tracers. Contributions of celestial gamma-ray events
are small, their variability is intrinsically different due to the
coded-mask modulation of the background model supplemented by
dithering, and thus is expected only to affect global
normalization, but not the variability with time.

We then orthogonalize the different background model components
for an improved convergence of the fitting (also see \S 2.2 of
Diehl et al. 2006; this step ensures that each higher-order
background component carries new and independent information ).

%In Figure 2, we display examples of background-component
%spectra for the 1173 keV line case. Both two background spectra
%have shown significant line features around the 1773 keV line
%energy range, confirming independent contribution of each
%background component to the fitting.

Representing our data sets with background-models only, we obtain
reduced $\chi^2$ values of 1.185 and 1.194 for the 1173 and
1332~keV line bands (SE, 277824 {\em d.o.f}), and 0.665 and 0.663
for the 1173 and 1333~keV line bands (ME, 615038 {\em d.o.f.}),
respectively. Note that for the low number of counts in ME
spectra, $\chi^2$ statistics does not apply. Therefore, in our
model fitting approach (\S 3.3), we use Poissonion statistics and
maximize  the Poisson-likelihood function.

\subsection{Model fitting and spectra}

We combine above background models with a spatial model for sky
emission (Figure 3) to fit our data, allowing for adjustments of
fit parameters for background and sky intensities. For this, we
use a maximum-likelihood method ({\em spimodfit}, for more details
of methods, see Strong et al. 2005) based on Poissonion
statistics. We this derive per energy bin, the fitted parameter
values with uncertainties, the covariance matrices, and the fitted
model components. The counts per energy bin, per detector, and per
pointing are fitted to the background model described in \S 3.2
and the assumed sky map of celestial emission (e.g., $^{26}$Al
distribution obtained by COMPTEL, see Figure 3) as convolved into
the domain of spectra per detector, energy bin, and pointing
through the pointing sequence and the instrumental response: \beq
D_{e,d,p}=\sum_{m,n} A_{e,d,p}^{m,n}\beta_s I^{m,n} +
\sum_{i=1}^3\beta_{b,i} B^i_{e,d,p} + \delta_{e,d,p}, \enq where
{\it e,d,p} are indices for data space dimensions: energy,
detector, pointing; {\it m,n} indices for the sky dimensions
(galactic longitude, latitude); $A$ is the instrument response
matrix, $I$ is the intensity per pixel on the sky. Coefficients
$\beta_s$ for the sky map intensity (constant in time) and
$\beta_{b,1}$, $\beta_{b,2}$, $\beta_{b,3}$ for three background
intensities are derived (see \S 3.2 and Figure 2; different
normalizations allowed for each camera configuration of 19/18/17
functional detector elements due to failure of two Ge detectors),
and the amplitude $\beta_s$ comprises the resultant spectra of the
signal from the sky. $\delta$ is the count residue after the
fitting. Generally, a good fit will lead to residuals being
statistically distributed around zero (see Figure 4).

Separate fits are made for each of the Fe lines and for SE and ME,
respectively. These produce four spectra of sky brightness
amplitudes per energy bin. If background models are adequate, the
continuum energy bands should show sky amplitudes statistically
consistent with zero sky brightness. Non-zero sky brightness
should show up in the form of lines with a shape conforming to the
instrumental line width, possibly broader if the celestial line
energy is broadened already (as had been discussed for $^{26}$Al,
see Diehl et al. 2006). Due to the low intensity of expected
$^{60}$Fe emission, we do not attempt to derive line shape
information at all, and rather derive the $^{60}$Fe line amplitude
by fitting a Gaussian with fixed instrumental width to the {\em
spimodfit} result spectra, determining line centroids and
amplitudes in this way. From strong nearby instrumental lines, we
determine that a Gaussian with a width of 2.76~keV is an adequate
model for narrow lines in the regime of the $^{60}$Fe lines; we
use this shape to determine $^{60}$Fe line parameters from the
{\em spimodfit} result spectra. For a consistent celestial signal
from $^{60}$Fe, the lines should be around the $^{60}$Fe decay
energies of 1173 and 1333 keV and equal (within errors) in
amplitudes for all four data sets.

%%%%%%%%%%%%%%%%%%%%%%%%%%%%%%%%%%%%%%%%%%%%%%%%%%%%%%%%%%%%%%%%%
\begin{figure*}
\centering
\includegraphics[angle=0,width=16cm]{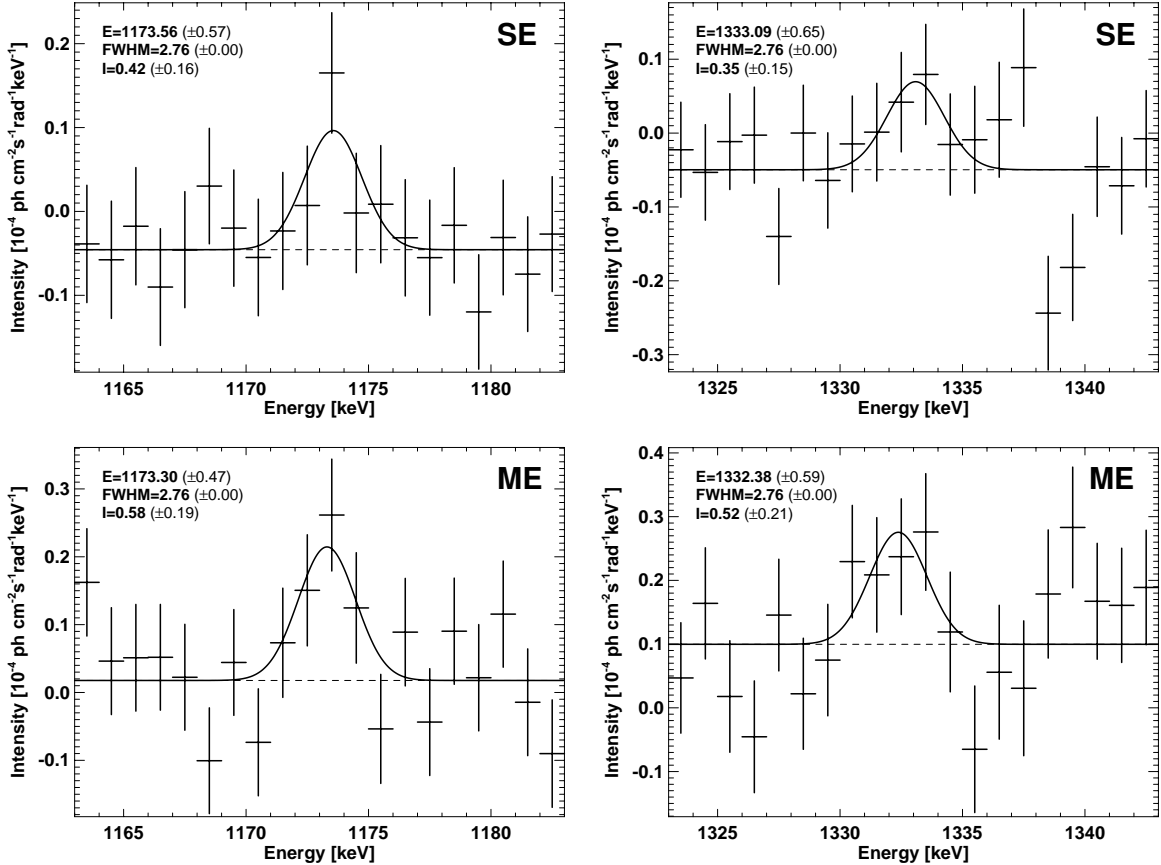}
\caption{The spectra of two gamma-ray lines of $^{60}$Fe from the
inner Galaxy: 1173 keV and 1333 keV. We have shown the results
both from SE and ME databases. The data points are fitted with
Gaussian profiles of fixed instrumental width (2.76 keV), and
fixed continuum slope (flat), with $\chi^2/d.o.f.$ of fits: 1.06
(SE) and 1.09 (ME) for the 1173 keV line and 1.16 (SE) and 1.11
(ME) for the 1333 keV line. For the SE database, we find a line
flux of $(4.2\pm 1.6)\times 10^{-5}{\rm ph\ cm^{-2}\ s^{-1}\
rad^{-1}}$ for the 1173 keV line and $ (3.5\pm 1.5)\times
10^{-5}{\rm ph\ cm^{-2}\ s^{-1}\ rad^{-1}}$ for the 1333 keV line.
For the ME database, the line flux is $ (5.8\pm 1.9)\times
10^{-5}{\rm ph\ cm^{-2}\ s^{-1}\ rad^{-1}}$ for the 1173 keV line
and $ (5.2\pm 2.1)\times 10^{-5}{\rm ph\ cm^{-2}\ s^{-1}\
rad^{-1}}$ for the 1333 keV line (also see Table 1).}
\end{figure*}
%%%%%%%%%%%%%%%%%%%%%%%%%%%%%%%%%%%%%%%%%%%%%%%%%%%%%%%%%%%%%%%%%
%%%%%%%%%%%%%%%%%%%%%%%%%%%%%%%%%%%%%%%%%%%%%%%%%%%%%%%%%%%%%%%%%
\begin{figure*}
\centering
\includegraphics[angle=0,width=10cm]{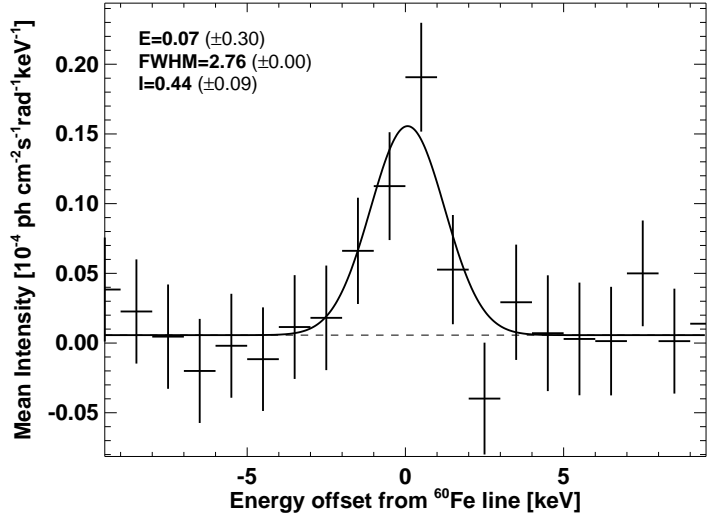}
\caption{The combined spectrum of the $^{60}$Fe signal in the
inner Galaxy, superimposing the four spectra of Figure 2. In the
laboratory, the line energies are 1173.23 and 1332.49 keV; here
superimposed bins are zero at 1173 and 1333 keV. We find a
detection significance of 4.9$\sigma$. The solid line represents a
fitted Gaussian profile of fixed instrumental width (2.76 keV),
and a flat continuum. The average line flux is estimated as $
(4.4\pm 0.9)\times 10^{-5}{\rm ph\ cm^{-2}\ s^{-1}\ rad^{-1}}$. }
\end{figure*}
%%%%%%%%%%%%%%%%%%%%%%%%%%%%%%%%%%%%%%%%%%%%%%%%%%%%%%%%%%%%%%%%%
%%%%%%%%%%%%%%%%%%%%%%%%%%%%%%%%%%%%%%%%%%%%%%%%%%%%%%%%%%%%%%%%%
\begin{figure*}
\centering \includegraphics[angle=0,width=15cm]{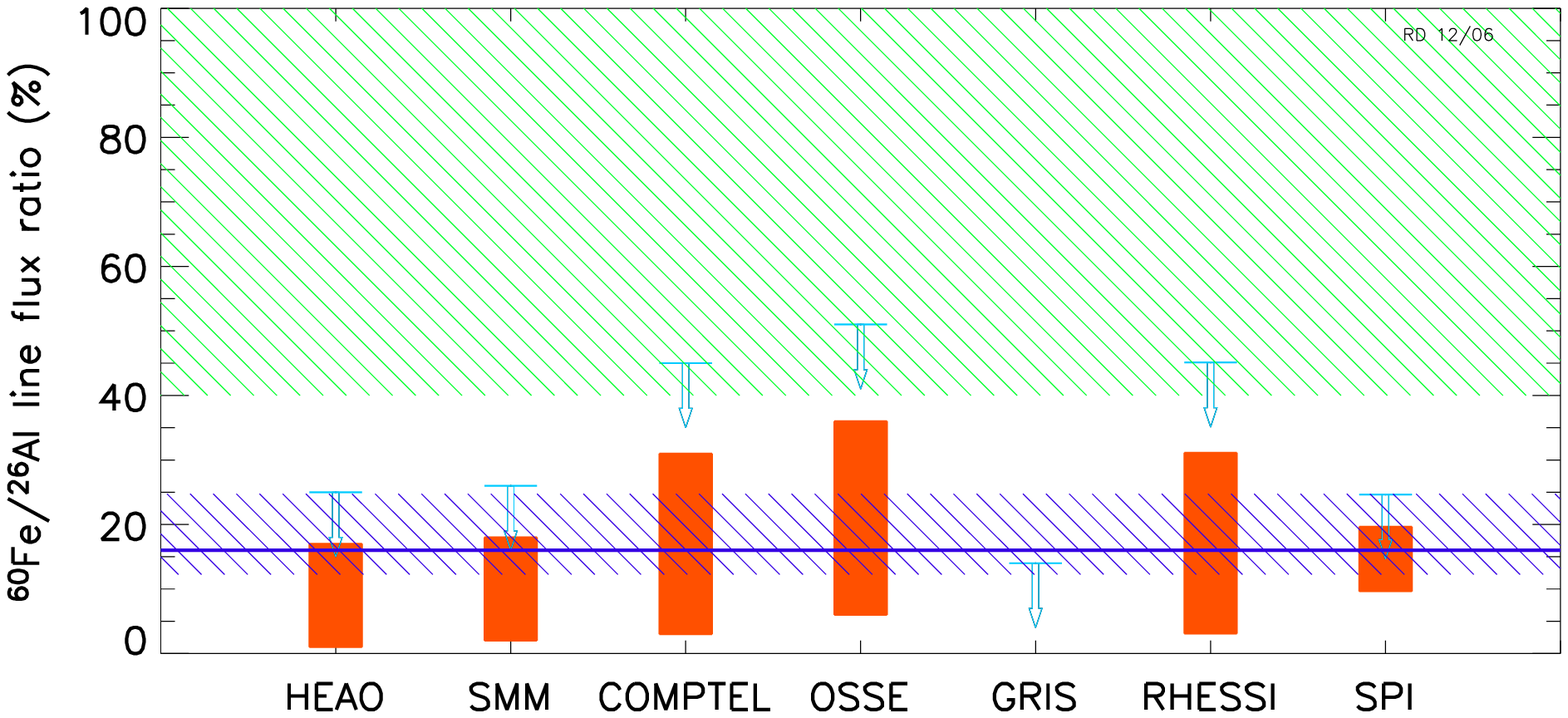}
\caption{Flux ratio of the gamma-ray lines from the two long-lived
radioactive isotopes $^{60}$Fe/$^{26}$Al from several
observations, including our SPI result (also see Table 2), with
upper limits shown at 2$\sigma$ for all reported values, and
comparison with the recent theoretical estimates (the upper
hatched region from Prantzos 2004; the horizontal line taken from
Timmes et al. 1995; the lower hatched region, see Limongi \&
Chieffi 2006). Our present work finds the line flux ratio to be $
(14.8\pm 6.0)\%$. Note that the primary instrument results on
$^{60}$Fe is compromised in the figure by different and uncertain
$^{26}$Al fluxes used for the $^{60}$Fe/$^{26}$Al ratio (see Table
2 for details).  }
\end{figure*}
%%%%%%%%%%%%%%%%%%%%%%%%%%%%%%%%%%%%%%%%%%%%%%%%%%%%%%%%%%%%%%%%%

\section{Results}

\subsection{SPI detection of $^{60}$Fe}

From independently fitting the SE and ME data sets at 1 keV-wide
energy bins in the energy ranges of the two lines of $^{60}$Fe
emission  (1173 keV and 1333 keV), we obtain the four spectra from
$spimodfit$ (\S 3.3) shown in Figure 5. Our fitted model consists
of the sky intensity distribution of $^{26}$Al from 9-year COMPTEL
observations (Figure 3, maximum entropy image as a standard input,
Pl\"uschke et al. 2001), together with the background model based
on measurements in energy bands adjacent to the $^{60}$Fe lines
and on background tracers (see above). We find excess amplitudes
for the celestial component in each of the four spectra.

Residual background imperfections are small except for the
instrumental line at 1333 keV. The continuum flux levels are close
to zero within $10^{-5} {\rm ph\ cm^{-2}\ s^{-1}\ rad^{-1}\
keV^{-1}}$, without apparent energy dependence, and adding a
linear slope to our spectral fit function does not improve the fit
\footnote{The diffuse gamma-ray continuum emission around
$^{60}$Fe line energies in the inner Galaxy is $\sim 2\times
10^{-6} {\rm ph\ cm^{-2}\ s^{-1}\ rad^{-1}\ keV^{-1}}$ from
COMPTEL measurements (Strong et al. 1999), which is well below the
error bars of spectral counts in Figure 5.  } . The $^{69}$Ge line
at 1337 keV has also been eliminated rather well, though not
completely. Suppression is a factor $\sim$ 4 better than our
previous work (Harris et al. 2005). However, this strong
instrumental line still leads to a spectral signal which is
recognized clearly as an artifact and thus not confused with the
celestial line.

We then determine the $^{60}$Fe line parameters in these spectra
fitting a continuum and Gaussians representing the SPI spectral
response (\S 2). The width of the Gaussian profiles has been fixed
at 2.76~keV, the value derived from simultaneously-measured
adjacent instrumental-line shapes (e.g. the instrumental line at
1107 keV). We thus obtain $^{60}$Fe line positions and intensities
from fitting the four spectra of Figure 5, where the total line
flux for each line in the inner Galaxy region (e.g. $-30^\circ
<l<30^\circ $ and $-10^\circ <b<10^\circ$) is determined from the
intensity of our fitted Gaussians and from the normalization of
the input sky map. We present line fluxes of each of the two
$^{60}$Fe lines for the SE and ME data in Table 1, respectively.
All line flux values are consistent within uncertainties.

A superposition of the four spectra of Fig. 5 is shown in Fig. 6.
Line energies of the $^{60}$Fe lines in the laboratory are 1173.23
and 1332.49 keV. For this superposition, we therefore define the
zero of the relative energy axis at 1173 and 1333 keV, to derive
the summed spectrum of all $^{60}$Fe signals. The line flux
estimated from the combined spectrum is $(4.4\pm 0.9)\times
10^{-5}{\rm ph\ cm^{-2}\ s^{-1}\ rad^{-1}}$. Our significance
estimate for the combined spectrum is $\sim 4.9\sigma$, adding
uncertainties of the individual spectra in quadrature. This
improves upon earlier $^{60}$Fe signal reports from RHESSI (Smith
2004) and the first year of SPI data (Harris et al. 2005).

The signal of celestial $^{60}$Fe is very weak, and marginally
significant ($< 3\sigma$) in each of our four spectra. Therefore
we cannot evaluate line shape information. The lines in our
spectra appear well represented by Gaussians with instrumental
widths (Fig. 5 and 6), suggesting that the celestial $^{60}$Fe
lines are intrinsically-narrow lines. If we assume a line
broadening of 1 keV, line fluxes would increase by $\sim 6\%$.

\subsection{$^{60}$Fe and $^{26}$Al}

It has been argued that $^{26}$Al and $^{60}$Fe share at least
some of the same production sites, i.e. massive stars and
supernovae (e.g., Timmes et al. 1995; Limongi \& Chieffi 2006). In
addition both are long-lived radioactive isotopes, so we have good
reasons to believe their gamma-ray distributions are similar as
well. Therefore we adopt the sky distribution of $^{26}$Al
gamma-rays as our model for celestial $^{60}$Fe gamma-ray
distribution.

% We have used the COMPTEL $^{26}$Al maximum entropy map as the
%standard sky model applied in the model fittings. While the
%maximum entropy map includes some structures and patches in the
%Galactic plane, we also used the COMPTEL $^{26}$Al MREM map
%(Figure 3, Pl\"uschke et al. 2001), a smoother one in model
%fittings.

From our model-fitting approach (\S 3.3), in principle, the
spectral result depends on the input sky map (also see Eq. 1). We
do not know the real distribution of $^{60}$Fe sources in the
Galaxy, but we need a realistic model for these sources to derive
correct spectra and line intensity.  As an alternative to the
COMPTEL $^{26}$Al maximum entropy and MREM sky maps, we also tried
a different sky map: an exponential-disk model with scale radius 4
kpc, and scale height 180 pc. This should also be a good
first-order representation of $^{26}$Al emission from the inner
Galaxy, and but avoids the fine structure inputs from the COMPTEL
image measurements which may partly arise from instrumental or
analysis imperfections. We find that different input sky maps do
not change the line profiles and intensities significantly
(variations of the line fluxes are within 10\% of the quoted
values, i.e. below the uncertainties). If we use very different
and scientifically implausible sky distribution models, e.g., a
point source at $l = \pm 20^\circ$ , a bulge model (a Gaussian
profile in the Galactic center), or a COMPTEL $^{26}$Al map zeroed
in the inner region of $-40^\circ < l <40^\circ,\ -10^\circ<b<
10^\circ$, then no signals are found near the $^{60}$Fe line
energies.

If we hypothesize a massive-star origin for 60Fe, then the
measurement of the gamma-ray flux ratio $^{60}$Fe/$^{26}$Al is
important for discussions of the astrophysical nucleosynthesis
origins of the two radioactive isotopes, and the nuclear physics
involved in models for their production (addressing the uncertain
nuclear reaction cross sections and half-lives). For this purpose,
we apply the same analysis method on the $^{26}$Al data from the
same observations, analyzing the 1785 --- 1826 keV energy band
(1~keV bins). We again generate a specific background model (from
adjacent energy bands and the GEDSAT background intensity tracer),
and apply the same input sky map (the 9-year COMPTEL $^{26}$Al
maximum entropy map) in model fitting the SE data. This yields a
flux ratio of $^{60}$Fe/ $^{26}$Al in a self-consistent way, with
$ F(^{60}{\rm Fe})/F(^{26}{\rm Al}) = (14.8 \pm 6.0)\%$. Here, the
uncertainty has been estimated from the respective model-fitting
uncertainties of the two SE databases. Alternatively, from the
combined spectrum of $^{60}$Fe lines (Fig. 6), and adopting the
$^{26}$Al intensity measured with SPI before ($F(^{26}{\rm Al})=
(3.04\pm 0.31)\times 10^{-4}{\rm ph\ cm^{-2}\ s^{-1}\ rad^{-1} }$
for the inner Galaxy, see Fig. 5 of Diehl et al. 2006), we obtain
$F(^{60}{\rm Fe})/F(^{26}{\rm Al})= (14.5 \pm 4.0) \%$.

\subsection{The Cygnus and Vela regions}

The Cygnus region is one of the most active nearby star formation
regions in our Galaxy. It contains a large number of massive stars
and rich OB associations at a distance of 1-2 kpc (e.g.
Kn\"odlseder 2000; Le Duigou \& Kn\"odlseder 2002), and has been
recognized as a prominent source of $^{26}$Al with COMPTEL (Diehl
et al. 1995a, Pl\"uschke et al. 2001) and with {\em INTEGRAL}/SPI
(Kn\"odlseder et al. 2004). From these massive stars, $^{60}$Fe
gamma-rays at the 1173 keV and 1333 keV decay lines may also be
expected. Since the majority of Cygnus region star-clusters are
young ($\sim 3$ Myr), from population synthesis studies of the
massive stars in the Cygnus region it has been suggested that the
$^{60}$Fe production is low, consistent with the small number of
recent supernova events inferred for this region ($F(1173\ {\rm
keV})\sim 2\times 10^{-6}{\rm ph\ cm^{-2}\ s^{-1}}$, see
Kn\"odlseder et al. 2002).

The Vela region in the southern sky includes even more nearby
massive stars, and prominent candidate $^{26}$Al sources such as
the Wolf-Rayet star WR 11 in the binary system $\gamma^2$Vel at
$d\sim 260$ pc (van der Hucht et al. 1997), and the core-collapse
supernova remnant Vela SNR at $d\sim 250$ pc (Cha et al. 1999).
1809 keV line emission of $^{26}$Al from the Vela region has been
reported from COMPTEL (Diehl et al. 1995b). And this emission is
not attributed to the nearby $\gamma^2$Vel source (Oberlack et al.
2000) and may be either diffuse or due to supernovae in the
region. Therefore the Vela region should be a good candidate for
gamma-ray line emission from $^{60}$Fe.

Splitting our model sky map into independent components to fit the
data, we do not see significant contributions from either region;
our estimated upper limits for $^{60}$Fe gamma-rays from the
Cygnus ($68^\circ <l<92^\circ$, $-10^\circ < b<14^\circ$) and Vela
($255^\circ <l<273^\circ$, $-9^\circ < b<9^\circ$) regions are
$\sim 1.1\times 10^{-5}{\rm ph\ cm^{-2}\ s^{-1}}$ ($2\sigma$).
More detailed studies are in progress (Schanne et al. 2007, and
Martin et al., in preparation).

\section{Summary and discussions}

We report the detection of $^{60}$Fe decay gamma-ray lines in the
Galaxy from 2.5 years of SPI observations. With our new
measurements we detect both the 1173 keV and 1333 keV line of
$^{60}$Fe from SPI single and multiple-detector events. Combining
our four spectra from independent model fits we obtain an
$^{60}$Fe signal from the Galaxy with a significance of $4.9
\sigma$. This improves upon our previous measurements (see Harris
et al. 2005). The average $^{60}$Fe line flux from the inner
Galaxy region is $ (4.4\pm 0.9)\times 10^{-5}{\rm ph\ cm^{-2}\
s^{-1}\ rad^{-1}}$, assuming intrinsically narrow lines and a sky
distribution equal to that of $^{26}$Al as measured by COMPTEL
(Pl\"uschke et al. 2001). From the same observations and analysis
procedure applied to $^{26}$Al, we find a flux ratio of $^{60}$Fe/
$^{26}$Al of $(14.8 \pm 6.0) \%$. The $^{60}$Fe signals are too
weak to determine line shape details; it appears that a Gaussian
with the width of the instrumental resolution can fit the data
well. This would imply that broadening of $^{60}$Fe lines from
astrophysical processes is not significant; most of the $^{60}$Fe
may be distributed in a rather normal interstellar medium with
turbulent velocities below $\sim$300~km~s$^{-1}$. In order to
investigate the variations over the Galaxy, we search for
$^{60}$Fe emission from the Cygnus and Vela regions, and do not
find $^{60}$Fe signals.

%%%%%%%%%%%%%%%%%%%%%%%%%%%%%%%%%%%%%%%%%%%%%%%%%%%%%%%%%%%
Many experiments and efforts were made (see Table 2 and Fig. 7) to
measure the $^{60}$Fe gamma-ray emission that was predicted by
theory -- we now provide the most significant detection to date.
In Fig. 7, we show the previous constraints on the flux ratio of
$^{60}$Fe/ $^{26}$Al together with the result of this work, and
compare the observational results with different theoretical
predictions. The earliest observational limit was given from
HEAO-3, $F(^{60}{\rm Fe})/F(^{26}{\rm Al}) = 0.09\pm 0.08$, an
upper limit being 0.27 (Mahoney et al. 1982) (in Fig. 7, we chose
to give limits at 2$\sigma$ for all reported values below a
significance of 3$\sigma$). Another limit was obtained with the
SMM Gamma-Ray Spectrometer, a flux ratio of $ 0.1\pm 0.08$, the
upper limit being $\sim 0.27$ (Leising \& Share 1994). OSSE aboard
the COMPTON Observatory gave a flux ratio of $ 0.21 \pm 0.15$, and
the upper limit is $\sim 0.51$ (Harris et al. 1994, 1997). COMPTEL
aboard the COMPTON Observatory also found $^{60}$Fe gamma-rays,
and reported a flux ratio value of $^{60}$Fe/ $^{26}$Al of $
0.17\pm 0.135$, which translates into an upper limit $\sim 0.44$
(Diehl et al. 1997). The Gamma-Ray Imaging Spectrometer (GRIS)
reported an upper limit for the ratio of $< 0.14$ (2$\sigma$, Naya
et al. 1998). RHESSI then reported the first detection of
$^{60}$Fe gamma-ray lines, and gave a flux ratio $0.17\pm 0.13$
with two-year data (Smith 2004). The first year data of SPI gave a
flux ratio $ 0.11\pm 0.07$ (Harris et al. 2005), and the present
analysis of the 2.5-year SPI data finds a flux ratio $0.148\pm
0.06$.

Theoretical predictions of the ratio of $^{60}$Fe/ $^{26}$Al have
undergone some changes since Timmes et al. (1995) published the
first detailed theoretical prediction. In their paper, they
combine a model for $^{26}$Al and $^{60}$Fe nucleosynthesis in
supernova explosions with a model of chemical evolution, to
predict that the steady production rates are $(2.0\pm 1.0)
M_\odot\ {\rm Myr^{-1}}$ for $^{26}$Al, and $(0.75\pm 0.4)
M_\odot\ {\rm Myr^{-1}}$ for $^{60}$Fe, which corresponds to a
gamma-ray flux ratio $F(^{60}{\rm Fe})/F(^{26}{\rm Al})= 0.16\pm
0.12$. This prediction would be consistent with our present
measurements. Since 2002, theoreticians have improved various
aspects of the stellar-evolution models, including improved
stellar wind models and the corresponding mass loss effects on
stellar structure and evolution, of mixing effects from rotation,
and also updated nuclear cross sections in the nucleosynthesis
parts of the models. As a result, predicted flux ratios $^{60}$Fe/
$^{26}$Al rather fell into the range $ 0.8\pm 0.4$ (see Prantzos
2004, based on, e.g. Rauscher et al. 2002, Limongi \& Chieffi
2003) -- such high values would be inconsistent with several
observational limits and our SPI result (but see Woosley \& Heger
2007 for comments on nuclear reaction rate updates ). Recently,
new calculations of $^{26}$Al and $^{60}$Fe yields from massive
stars of solar metallicity have been presented (Limongi \& Chieffi
2006; Heger \& Woosley 2007). Limongi \& Chieffi (2006) combined
their individual yields, using a standard stellar-mass
distribution function, to produce an estimate of the $^{60}$Fe/
$^{26}$Al gamma-ray flux ratio expected from massive stars. Their
calculations yield again a lower prediction for the $^{60}$Fe/
$^{26}$Al flux ratio of $ 0.185\pm 0.0625$, which is again
consistent with the observational constraints (see Figure 7).

In summary, uncertainties still exist, both in models and
measurements of $^{60}$Fe. On the theory side, stellar evolution
in late stages is complex, nuclear reactions include neutron
capture on unstable Fe isotopes, and explosive nucleosynthesis
adds yet another complex ingredient. On the experimental side,
cosmic ray induced $^{60}$Co radioactivity in the instrument and
spacecraft and the limitations of spatial resolutions and
sensitivity are issues reflected in the substantial uncertainties
in experimental values. With more {\em INTEGRAL}/SPI data to come,
and also with the development of next-generation gamma-ray
spectrometers/telescopes, gamma-ray observations hopefully can
help with an independent view on the astrophysical model
components.

\section*{Acknowledgments}
We are grateful to the referee for the constructive comments. The
{\em INTEGRAL} project is supported by government grants in all
member states of the hardware teams. The SPI project has been
completed under responsibility and leadership of CNES. We are
grateful to ASI, CEA, CNES, DLR, ESA, INTA, NASA, and OSTC for
support.

{}

\begin{table}
\caption{$^{60}$Fe intensity in the inner Galaxy} \label{table:1}
\begin{center}
\begin{tabular}{l l}
\hline \hline  &  Flux ($10^{-5}\ {\rm ph\ cm^{-2}\ s^{-1}\
rad^{-1}}$)
  \\
\hline 1173 keV (SE) & $4.2\pm 1.6$  \\
1173 keV (ME) & $5.8\pm 1.9$  \\
1333 keV (SE) & $3.5\pm 1.5$ \\
1333 keV (ME) & $5.2\pm 2.1$ \\

\hline
\end{tabular}
\end{center}
\end{table}

\begin{table}
\caption{Different measurements of $^{60}$Fe flux in units of
$10^{-5}$ ph cm$^{-2}$ s$^{-1}$ rad$^{-1}$ from the inner Galaxy
and $^{60}$Fe/$^{26}$Al flux ratio } \label{table:2}
\begin{center}
\begin{tabular}{l c c l}
\hline \hline Experiments & $^{60}$Fe flux & $^{60}{\rm Fe}/^{26}{\rm Al}$ & references  \\
\hline

HEAO-3 & $5.3\pm 4.3$ & $0.09\pm 0.08$  & Mahoney et al. 1982 \\
SMM & $2.9\pm 2.5$ & $0.1\pm 0.08$    &  Leising \& Share 1994 \\
OSSE & $6.3\pm 4.5$  &$0.21 \pm 0.15$    &  Harris et al. 1997 \\
COMPTEL &  $6.7\pm 5.4$    &  $0.17\pm 0.135$     & Diehl et al. 1997  \\
GRIS & $<6.8 (2\sigma)$ & $<0.14 (2\sigma)$  & Naya et al. 1998 \\
RHESSI &  $3.6\pm 1.4$  & $0.17\pm 0.13$ & Smith 2004  \\
SPI &  $4.4\pm 0.9$ &  $0.148\pm 0.06$ & this work \\
\hline
\end{tabular}
\end{center}
\end{table}

\end{document}